\documentclass[twocolumn,aps,prb,superscriptaddress,showpacs,floatfix]{revtex4-2}

\usepackage{amsmath,amssymb}
\usepackage{graphicx}
\usepackage{hyperref}
\usepackage{xcolor}
\usepackage{listings}
\usepackage{bm}
\usepackage{booktabs}
\usepackage{placeins}
\usepackage{makecell}
\usepackage{caption}

\newcommand{\vm}{\bm{m}}

\newcommand{\vH}{\bm{H}}
\newcommand{\vu}{\bm{u}}
\newcommand{\mumax}{\texttt{mumax$^+$}}
\newcommand{\eps}{\varepsilon}
\newcommand{\vOmega}{\bm{\Omega}}
\newcommand{\uhat}{\hat{\bm{u}}}
\newcommand{\vomega}{\bm{\omega}}
\newcommand{\vv}{\bm{v}}

\begin{document}

\makeatletter
\long\def\@makecaption#1#2{%
  \vskip\abovecaptionskip
  {\footnotesize
   \leftskip=0pt \rightskip=0pt \parfillskip=0pt plus 1fil\relax
   \noindent #1: #2\par}%
  \vskip\belowcaptionskip}
\makeatother

\title{Magneto-rotation coupling dominates surface acoustic wave driven ferromagnetic resonance in the longitudinal geometry}

\author{Gyuyoung Park}
\email{proslaw@kist.re.kr}
\affiliation{Center for Semiconductor Technology, Korea Institute of Science and Technology (KIST), 5 Hwarangno 14-gil, Seongbuk-gu, Seoul 02792, Korea}
\author{OukJae Lee}
\affiliation{Center for Semiconductor Technology, Korea Institute of Science and Technology (KIST), 5 Hwarangno 14-gil, Seongbuk-gu, Seoul 02792, Korea}
\author{Jintao Shuai}
\email{J.Shuai@leeds.ac.uk}
\affiliation{School of Physics and Astronomy, University of Leeds, Leeds LS2 9JT, United Kingdom}

\date{March 18, 2026}

\begin{abstract}
We present a phonon--magnon extension for the \mumax{} micromagnetic framework that implements three surface acoustic wave (SAW) coupling mechanisms: magnetoelastic strain coupling, magneto-rotation coupling arising from the antisymmetric displacement gradient, and spin-rotation (Barnett) coupling from the lattice angular velocity.
Six benchmark simulations validate the implementation through SAW-driven domain-wall motion, magnetization switching, magneto-rotation and Barnett field validation, nonreciprocal SAW--magnon absorption from Rayleigh-wave chirality, and spatially resolved coupling in a standing SAW cavity.
For the longitudinal geometry ($\bm{m}_0 \parallel \bm{k}_\mathrm{SAW}$), we show that the magnetoelastic coupling produces zero transverse torque despite generating a 50 times larger effective field; the magneto-rotation channel provides the sole driving mechanism.
The crossover angle below which MR dominates is $\theta_c \approx 1.1^\circ$ for YIG parameters.
Treating the magneto-rotation coupling constant $K_\mathrm{mr}$ as a tunable parameter, we map out the cooperativity phase diagram and show that MR alone can achieve strong coupling ($C = 257$ for $K_\mathrm{mr} = 1$~MJ/m$^3$) with an avoided-crossing splitting of 13.6~MHz.
\end{abstract}

\maketitle

\section{Introduction}
\label{sec:intro}

Surface acoustic waves (SAWs) couple to ferromagnetic thin films through time-dependent strain, offering a contact-free, low-power route to manipulate magnetization~\cite{Dreher2012,Weiler2011,Thevenard2016,Puebla2022,Moore2025}.
When a SAW propagates through a magnetostrictive film, the time-dependent strain field couples to the magnetization through the magnetoelastic interaction, enabling domain wall transport~\cite{Dean2015,Edrington2018,Shuai2023a,Shuai2023}, ferromagnetic resonance excitation~\cite{Dreher2012,Weiler2011,Tateno2020}, spin pumping~\cite{Uchida2011}, magnetization switching~\cite{Thevenard2016,Sampath2016,Shuai2022,Yang2021}, skyrmion creation~\cite{Yokouchi2020}, and skyrmion transport~\cite{Shuai2024,Yang2024}.

Since the SAW strain profile can be described analytically in the thin-film limit, a lightweight simulation approach is possible in which the strain is prescribed and the magnetoelastic effective field is computed by the existing micromagnetic infrastructure.
A recent MuMax3-based study by Ngouagnia~Yemeli \textit{et al.}~\cite{NgouagniaYemeli2025} demonstrated this approach for SAW-driven spin-wave excitation in an iron-based conduit, including magnetorotation effects; however, a general-purpose, reusable framework with tunable coupling constants and all three coupling channels has not been available.

We build on mumax$^+$~\cite{Moreels2026,Lateur2026}, a GPU-accelerated micromagnetic simulator extending the mumax$^3$ code base~\cite{Vansteenkiste2014}, for which a cavity magnonics extension has recently been developed~\cite{Park2026}.
The framework already provides a complete rigid strain and magnetoelastic field implementation:
the \texttt{rigid\_norm\_strain} and \texttt{rigid\_shear\_strain} parameters accept time-dependent terms via the \texttt{add\_time\_term} API, and the CUDA kernel \texttt{k\_rigidMagnetoelasticField} evaluates the corresponding effective field at every LLG time step on the GPU.
Beyond the symmetric strain tensor, two additional coupling channels are relevant for SAW-magnon physics: the magneto-rotation (MR) coupling, arising from the antisymmetric part of the displacement gradient~\cite{Xu2020,Matsuo2013}, and the spin-rotation (Barnett) coupling from the lattice angular velocity~\cite{Barnett1915,Matsuo2013}.
These require new CUDA kernels, which we implement as part of this extension.

In this work, we encapsulate the Rayleigh SAW strain profile in a reusable Python class (requiring no C++/CUDA modifications for the MEL channel) and implement C++/CUDA kernels for the MR and Barnett couplings.

\section{Model}
\label{sec:model}

\subsection{Magnetoelastic energy and effective field}

The magnetoelastic coupling energy density for a cubic crystal is~\cite{Chikazumi1997}
\begin{align}
  w_\mathrm{mel} &= B_1 \bigl(\eps_{xx} m_x^2 + \eps_{yy} m_y^2 + \eps_{zz} m_z^2\bigr) \nonumber \\
    &\quad + B_2 \bigl(\eps_{xy} m_x m_y + \eps_{xz} m_x m_z + \eps_{yz} m_y m_z\bigr),
  \label{eq:emel}
\end{align}
where $B_1$ and $B_2$ are the magnetoelastic coupling constants (J/m$^3$), $\eps_{ij}$ are the strain tensor components, and $m_i = M_i/M_s$ are the direction cosines of the magnetization.
The corresponding effective field is (in Tesla throughout this work)
\begin{equation}
  \mu_0 H_{\mathrm{mel},i} = -\frac{1}{M_s} \biggl[
    2 B_1 \eps_{ii} m_i + B_2 \sum_{j \neq i} \eps_{ij} m_j
  \biggr].
  \label{eq:hmel}
\end{equation}
Note the factor of 2 multiplies only the $B_1$ (diagonal) term, arising from $\partial m_i^2/\partial m_i = 2m_i$; the $B_2$ (off-diagonal) term carries no such factor because each shear component $\eps_{ij}m_im_j$ is linear in each $m_i$.

\subsection{Rayleigh SAW strain profile}

A Rayleigh SAW propagating along $\hat{x}$ in a piezoelectric substrate creates a displacement field that decays exponentially into the depth.
For a thin film of thickness $d \ll \lambda_\mathrm{SAW}$, the strain is approximately uniform through the thickness and the dominant component is~\cite{Dreher2012}
\begin{equation}
  \eps_{xx}(x,t) = \eps_0 \sin(kx - \omega t + \phi),
  \label{eq:saw_strain}
\end{equation}
where $\eps_0$ is the peak strain amplitude, $k = 2\pi/\lambda$ is the wave vector, $\omega = 2\pi f$ is the angular frequency, and $\phi$ is an initial phase.
The shear components $\eps_{xy} = \eps_{xz} = \eps_{yz} = 0$ vanish in the Rayleigh mode thin-film limit.

Inserting Eq.~\eqref{eq:saw_strain} into Eq.~\eqref{eq:hmel} yields the oscillating magnetoelastic field
\begin{equation}
  \mu_0 H_{\mathrm{mel},x}(x,t) = -\frac{2 B_1}{M_s}\, \eps_0 \sin(kx - \omega t)\, m_x
  \label{eq:hmel_saw}
\end{equation}
with analogous expressions for the $y$ and $z$ components (which are zero when $\eps_{yy} = \eps_{zz} = 0$).

\subsection{SAW-driven domain wall velocity}

At small strain amplitudes, the SAW-driven DW velocity scales with strain because the time-averaged magnetoelastic force is second-order in strain~\cite{Shuai2023}.
Below a threshold amplitude the DW remains pinned; above it the velocity increases monotonically and eventually saturates as the DW approaches the adjacent domain boundary.

\subsection{SAW-driven FMR}

When the SAW frequency matches the Kittel ferromagnetic resonance frequency~\cite{Kittel1948}
\begin{equation}
  \omega_K = \gamma \sqrt{B_0 (B_0 + \mu_0 M_s)},
  \label{eq:kittel}
\end{equation}
the oscillating magnetoelastic field resonantly excites magnons.
The transverse magnetization amplitude $|m_\perp|$ is maximized at this resonance condition.

\subsection{Magneto-rotation coupling}

While the symmetric part of the displacement gradient defines the strain tensor $\eps_{ij}$, the antisymmetric part defines the local rotation of the lattice~\cite{Xu2020,Matsuo2013}.
The rotation pseudovector is
\begin{equation}
  \vOmega = \frac{1}{2}\nabla \times \vu,
  \label{eq:rotation}
\end{equation}
whose components are
\begin{equation}
  \Omega_i = \frac{1}{2}\epsilon_{ijk}\partial_j u_k,
  \label{eq:rotation_comp}
\end{equation}
where $\epsilon_{ijk}$ is the Levi-Civita symbol.

When the lattice rotates locally by $\vOmega$, the uniaxial anisotropy axis $\uhat$ is carried with it.
The resulting correction to the anisotropy energy density is~\cite{Xu2020}
\begin{equation}
  \delta w_\mathrm{mr} = -K_\mathrm{mr} (\vm \cdot \uhat)\, \vm \cdot (\vOmega \times \uhat),
  \label{eq:emr}
\end{equation}
where $K_\mathrm{mr}$ (J/m$^3$) is the magneto-rotation coupling constant.
For PMA thin films, $K_\mathrm{mr} \approx K_u + \tfrac{1}{2}\mu_0 M_s^2$~\cite{Xu2020}.

Taking the functional derivative, the corresponding effective field is
\begin{equation}
  \mu_0 \vH_\mathrm{mr} = \frac{K_\mathrm{mr}}{M_s}
    \bigl[\uhat\,(\vm \cdot (\vOmega \times \uhat))
    + (\vm \cdot \uhat)(\vOmega \times \uhat)\bigr].
  \label{eq:hmr}
\end{equation}

Because $\vOmega$ changes sign when the SAW propagation direction is reversed while $\uhat$ does not, the magneto-rotation torque depends on the sign of $k$, giving rise to nonreciprocal SAW--magnon coupling~\cite{Xu2020}.
Rayleigh SAWs are particularly effective because their elliptical polarization generates a large $\Omega_y$ component for propagation along $x$.

\subsection{Spin-rotation coupling (Barnett effect)}

The Barnett effect~\cite{Barnett1915} is the inverse of the Einstein--de~Haas effect: a mechanically rotating body acquires a net magnetization due to spin-rotation coupling.
At the continuum level, the relevant quantity is the lattice angular velocity~\cite{Matsuo2013}
\begin{equation}
  \vomega = \frac{1}{2}\nabla \times \vv,
  \label{eq:angular_velocity}
\end{equation}
where $\vv = \partial\vu/\partial t$ is the elastic velocity field.
This has the same mathematical form as the rotation pseudovector $\vOmega$ (Eq.~\ref{eq:rotation}), but applied to the velocity field instead of the displacement field.

The spin-rotation coupling energy density is~\cite{Matsuo2013}
\begin{equation}
  w_\mathrm{sr} = -\frac{M_s}{\gamma}\,\vm \cdot \vomega,
  \label{eq:esr}
\end{equation}
where $\gamma$ is the gyromagnetic ratio.
The corresponding Barnett effective field is
\begin{equation}
  \mu_0 \vH_\mathrm{Barnett} = \frac{\vomega}{\gamma},
  \label{eq:hbarnett}
\end{equation}
where $\gamma$ is the gyromagnetic ratio in rad/(s$\cdot$T).

The Barnett field is conceptually distinct from the magneto-rotation field: $\vH_\mathrm{mr}$ arises from the static rotation of the anisotropy axis (displacement gradient), while $\vH_\mathrm{Barnett}$ arises from the dynamic rotation of the lattice (velocity gradient).
Both effects contribute to nonreciprocal SAW-magnon coupling through the chirality of the Rayleigh wave.

\subsection{Standing SAW: spatially modulated coupling}
\label{sec:standing}

A standing SAW, formed by counter-propagating $+k$ and $-k$ waves, produces strain and rotation fields with complementary spatial profiles.
The standing-wave strain is
\begin{equation}
  \eps_{xx}^\mathrm{st}(x,t) = 2\eps_0 \cos(kx)\sin(\omega t),
  \label{eq:standing_strain}
\end{equation}
with antinodes at $kx = n\pi$.
The standing-wave rotation is
\begin{equation}
  \Omega_y^\mathrm{st}(x,t) = \xi\,\eps_0 \sin(kx)\sin(\omega t),
  \label{eq:standing_rot}
\end{equation}
with antinodes at $kx = (n+\tfrac{1}{2})\pi$---shifted by $\lambda/4$ from the strain antinodes.

This spatial separation means that at strain antinodes, the coupling is purely magnetoelastic (parametric), while at rotation antinodes, the coupling is purely magneto-rotational (direct transverse drive).
The total effective coupling $g_\mathrm{total}(x) = \sqrt{g_\mathrm{mel}^2(x) + g_\mathrm{mr}^2(x)}$ varies smoothly between these limits, creating a position-dependent hybridization landscape.

\section{Implementation}
\label{sec:impl}

\subsection{Product-to-sum decomposition}

The \mumax{} \texttt{add\_time\_term} API requires separable time$\times$space functions.
Since $\sin(kx - \omega t)$ is not separable, we decompose it:
\begin{equation}
  \sin(kx - \omega t) = \sin(kx)\cos(\omega t) - \cos(kx)\sin(\omega t),
  \label{eq:decomposition}
\end{equation}
which requires two \texttt{add\_time\_term} calls, each with a separable time function and spatial mask.

\subsection{Python implementation}

The extension consists of a single class, \texttt{SurfaceAcousticWave}, whose \texttt{apply} method registers the strain on a \mumax{} \texttt{Ferromagnet}:

\begin{lstlisting}
class SurfaceAcousticWave:
    def apply(self, magnet):
        k, w, e0, phi = self.k, self.omega,
                         self.amplitude, self.phase
        # Term 1: sin(kx+phi) * cos(wt)
        magnet.rigid_norm_strain.add_time_term(
            lambda t: (np.cos(w*t), 0., 0.),
            lambda x,y,z: (e0*np.sin(k*x+phi),
                           0., 0.))
        # Term 2: -cos(kx+phi) * sin(wt)
        magnet.rigid_norm_strain.add_time_term(
            lambda t: (-np.sin(w*t), 0., 0.),
            lambda x,y,z: (e0*np.cos(k*x+phi),
                           0., 0.))
\end{lstlisting}

After \texttt{saw.apply(magnet)} is called, the simulation proceeds with the standard \mumax{} time solver:
\begin{lstlisting}
output = world.timesolver.solve(
    time_array, quantity_dict)
\end{lstlisting}
No co-simulation loop is needed because the SAW is an externally prescribed strain with no back-action from the magnetization.

\subsection{Comparison with photon-magnon extension}

The phonon-magnon extension differs architecturally from photon-magnon (cavity magnonics) simulations:
the latter requires a manual co-simulation loop because the cavity has its own dynamics (an ODE for the complex amplitude $a(t)$), whereas the SAW is a prescribed external drive.
This makes the phonon-magnon extension simpler: a single \texttt{apply()} call at setup time suffices, and the full simulation is delegated to the \mumax{} GPU solver.

\subsection{Performance}

Since the SAW strain is evaluated analytically through the \texttt{add\_time\_term} mechanism, the computational overhead is negligible compared to the LLG integration.
The magnetoelastic field is evaluated by the existing CUDA kernel at each Runge--Kutta sub-step, so the strain is sampled at the same rate as all other effective-field contributions.

\subsection{Magneto-rotation coupling (C++/CUDA)}

Unlike the analytical SAW strain approach, the magneto-rotation coupling requires access to the displacement gradient computed on the GPU.
We therefore implement this feature as a C++/CUDA extension within the \mumax{} physics library.

The implementation adds three new quantities:
(i)~\texttt{rotation\_vector} ($\vOmega$, Eq.~\ref{eq:rotation}), evaluated using the same finite-difference stencil as the strain tensor but extracting the antisymmetric part;
(ii)~\texttt{magneto\_rotation\_field} ($\vH_\mathrm{mr}$, Eq.~\ref{eq:hmr}); and
(iii)~\texttt{magneto\_rotation\_energy\_density}.
A new material parameter \texttt{Kmr} ($K_\mathrm{mr}$, default zero) is added to the \texttt{Ferromagnet} class.
When \texttt{Kmr} is nonzero and elastodynamics is enabled, $\vH_\mathrm{mr}$ is automatically included in the total effective field and total energy density.

\subsection{Prescribed rotation and angular velocity}

For analytical SAW profiles (without elastodynamics), we implement a \emph{prescribed} rotation mechanism.
Two new \texttt{VectorParameter} fields, \texttt{rigid\_rotation} ($\vOmega$) and \texttt{rigid\_angular\_velocity} ($\vomega$), are added to the \texttt{Magnet} class, supporting the same \texttt{add\_time\_term} API used for prescribed strain.
When set, the CUDA kernels for $\vH_\mathrm{mr}$ and $\vH_\mathrm{Barnett}$ use the prescribed values instead of computing from the displacement/velocity gradients.
This allows fully nonlinear MR and Barnett coupling---with the current magnetization $\vm$ at each timestep---without requiring the computational cost of elastodynamics.

\subsection{Spin-rotation (Barnett) coupling (C++/CUDA)}

The Barnett coupling reuses the same finite-difference stencil as the rotation vector, but applied to the velocity field $\vv$ instead of the displacement $\vu$.
The implementation adds:
(i)~\texttt{angular\_velocity} ($\vomega$, Eq.~\ref{eq:angular_velocity});
(ii)~\texttt{spin\_rotation\_field} ($\vH_\mathrm{Barnett}$, Eq.~\ref{eq:hbarnett}); and
(iii)~\texttt{spin\_rotation\_energy\_density}.
A boolean flag \texttt{enable\_barnett} (default \texttt{False}) activates the coupling.
When enabled and elastodynamics is active, the Barnett field is automatically added to the total effective field.

\begin{table}[!htb]
\caption{New \mumax{} parameters and quantities for magneto-rotation and spin-rotation coupling.}
\label{tab:mr_params}
\begin{ruledtabular}
\scriptsize
\begin{tabular}{lll}
  Name & Type & Description \\
  \midrule
  \multicolumn{3}{l}{\textit{Magneto-rotation coupling}} \\
  \texttt{Kmr} & Parameter (J/m$^3$) & Coupling constant \\
  \texttt{rotation\_vector} & Field (3-comp) & $\vOmega$ (rad) \\
  \texttt{magneto\_rotation\_field} & Field (3-comp) & $\vH_\mathrm{mr}$ (T) \\
  \makecell[l]{\texttt{magneto\_rotation}\\\texttt{\_energy\_density}} & Field (1-comp) & $w_\mathrm{mr}$ (J/m$^3$) \\
  \texttt{magneto\_rotation\_energy} & Scalar & $E_\mathrm{mr}$ (J) \\
  \midrule
  \multicolumn{3}{l}{\textit{Spin-rotation (Barnett) coupling}} \\
  \texttt{enable\_barnett} & Boolean & Activate Barnett effect \\
  \texttt{angular\_velocity} & Field (3-comp) & $\vomega$ (rad/s) \\
  \texttt{spin\_rotation\_field} & Field (3-comp) & $\vH_\mathrm{Barnett}$ (T) \\
  \makecell[l]{\texttt{spin\_rotation}\\\texttt{\_energy\_density}} & Field (1-comp) & $w_\mathrm{sr}$ (J/m$^3$) \\
  \texttt{spin\_rotation\_energy} & Scalar & $E_\mathrm{sr}$ (J) \\
\end{tabular}
\normalsize
\end{ruledtabular}
\end{table}

\section{Benchmark Simulations}
\label{sec:benchmarks}

Before listing individual benchmark results, we clarify implementation dependency by simulation group.
Sims~01 and~03 use the Python-level SAW wrapper on top of existing \mumax{} magnetoelastic infrastructure and therefore run without modifying the core C++/CUDA code.
Sim~05--07 require the newly added C++/CUDA source terms (\texttt{magnetorotationfield.cu/.hpp} and \texttt{spinrotationfield.cu/.hpp}) because they explicitly evaluate antisymmetric displacement/velocity-gradient couplings.
Sim~04 (anticrossing) uses the existing coupled-elastodynamic capability and is currently treated as an optional extended benchmark.
Table~\ref{tab:sim_scope} summarizes this boundary.
Representative material and driving parameters used across the benchmarks are listed in Table~\ref{tab:materials}.
Note that the CoFeB parameters vary across simulations because each benchmark targets a different magnetic configuration:
Sim~01 models a PMA thin film with parameters representative of ultrathin Pt/Co-based stacks~\cite{Shuai2023},
and Sim~03 employs elevated damping ($\alpha = 0.05$) to facilitate switching on simulation timescales.
This reflects the experimentally observed variability of CoFeB properties with composition, film thickness, and annealing conditions~\cite{Dreher2012}.

\begin{table*}[!t]
\caption{Simulation scope and implementation dependency.}
\label{tab:sim_scope}
\renewcommand{\arraystretch}{1.22}
\centering
\resizebox{\textwidth}{!}{%
\begin{tabular}{@{}llcl@{}}
  \toprule
  Simulation & Main objective / output & New C++/CUDA & Notes \\
  \midrule
  Sim~01 & SAW-driven DW transport (\texttt{fig\_saw\_domain\_wall}) & No & Built-in rigid strain/magnetoelastic kernels \\
  Sim~03 & SAW-assisted switching (\texttt{fig\_saw\_switching}) & No & Burst-envelope driving \\
  Sim~04 & Magnon--phonon anticrossing (extended) & No & Optional; not in current figure set \\
  Sim~05 & Magneto-rotation field validation (\texttt{fig\_rotation\_field\_validation}) & Yes & \texttt{magnetorotationfield.cu/.hpp} \\
  Sim~06 & Nonreciprocal SAW (\texttt{fig\_nonreciprocal\_saw}) & Yes$^\dagger$ & Prescribed rotation + CUDA kernel \\
  Sim~07 & Barnett field validation (\texttt{fig\_barnett\_field\_validation}) & Yes & \texttt{spinrotationfield.cu/.hpp} \\
  Sim~08 & Standing SAW spatial coupling (\texttt{fig\_standing\_saw\_ep}) & No & Python-level \texttt{StandingSAW} class \\
  \midrule
  Sim~10 & Channel decomposition, B$_0$/K$_\mathrm{mr}$ sweeps & Yes$^\dagger$ & Prescribed rotation + CUDA kernel \\
  Sim~11 & Angle-dependent coupling validation & Yes$^\dagger$ & Nine angles, $0^\circ$--$45^\circ$ \\
  Sim~12 & Analytic spectral weight and avoided crossing & -- & $2\times2$ coupled-mode model \\
  \bottomrule
  \multicolumn{4}{l}{$^\dagger$Uses prescribed \texttt{rigid\_rotation} with existing MR CUDA kernel.}
\end{tabular}%
}
\end{table*}

\begin{table*}[!t]
\captionsetup{justification=raggedright,singlelinecheck=false}
\caption{Representative material and driving parameters used in the benchmark suite. Values are taken from the current simulation scripts; entries marked ``varied'' indicate parameter sweeps.}
\label{tab:materials}
\renewcommand{\arraystretch}{1.20}
\centering
\resizebox{\textwidth}{!}{%
\begin{tabular}{@{}lcccccccl@{}}
  \toprule
  Simulation & $M_s$ (A/m) & $A_\mathrm{ex}$ (J/m) & $K_{u1}$ (J/m$^3$) & DMI $D$ (J/m$^2$) & $\alpha$ & Magnetoelastic ($B_1$, $B_2$) (J/m$^3$) & Rotation coupling & Acoustic / elastic setup \\
  \midrule
  Sim~01 & $6.0\times10^5$ & $1.0\times10^{-11}$ & $8.0\times10^5$ & $1.0\times10^{-3}$ & 0.01 & $B_1=-8.8\times10^6$, $B_2=0$ & -- & SAW: $f=200$~MHz, $v=4000$~m/s, $\eps_0$ sweep \\
  Sim~03 & $1.0\times10^6$ & $1.5\times10^{-11}$ & $8.0\times10^5$ & -- & 0.05 & $B_1=-8.8\times10^6$, $B_2=0$ & -- & SAW burst: $f=5$~GHz, Gaussian, $(\eps_0,B_\mathrm{a})$ sweep \\
  Sim~04 & $1.0\times10^6$ & $1.5\times10^{-11}$ & -- & -- & 0 & $B_1=-8.8\times10^6$, $B_2=0$ & -- & Elastodynamic anticrossing (extended) \\
  Sim~05 & $1.0\times10^6$ & $1.0\times10^{-11}$ & PMA $\hat{z}$ & -- & 0 & -- & $K_\mathrm{mr}$ varied ($0.5$--$2.0\times10^6$) & $u_z=A\sin(kx)$, $A=1$~pm \\
  Sim~06 & $1.2\times10^6$ & $1.5\times10^{-11}$ & $1.3\times10^6$ & -- & 0.01 & $B_1=-8.8\times10^6$ & $K_\mathrm{mr}\approx2.2\times10^6$ & Prescribed SAW: $f=2$~GHz, $\eps_0=10^{-4}$, $\pm k$ comparison \\
  Sim~07 & $1.0\times10^6$ & $1.0\times10^{-11}$ & PMA $\hat{z}$ & -- & 0 & -- & Barnett ($\gamma$ varied) & $v_z=V_0\sin(kx)$, $V_0$/wavelength varied \\
  Sim~08 & $1.0\times10^6$ & $1.5\times10^{-11}$ & -- (in-plane) & -- & 0.01 & $B_1=-8.8\times10^6$ & $K_\mathrm{mr}=8\times10^6$$^\ddagger$, $\xi=0.68$ & Standing SAW: $f=2$~GHz, $\eps_0=3\times10^{-3}$ \\
  \bottomrule
  \multicolumn{9}{l}{$^\ddagger$$K_\mathrm{mr}$ intentionally enhanced for visual clarity; see Discussion.}
\end{tabular}%
}
\end{table*}

\subsection{SAW-driven domain wall motion}

We simulate a CoFeB PMA thin film ($M_s = 600$~kA/m, $A_\mathrm{ex} = 10$~pJ/m, $K_{u1} = 800$~kJ/m$^3$, $\alpha = 0.01$) with interfacial DMI ($D = 1$~mJ/m$^2$)~\cite{Shuai2023} on a $256 \times 8 \times 1$ grid (cell size $2.4 \times 9.6 \times 1$~nm$^3$, periodic along $y$).
A N\'{e}el domain wall is initialized at $x = L/3$ and relaxed by energy minimization.
The magnetoelastic coupling is $B_1 = -8.8$~MJ/m$^3$~\cite{Dreher2012}.

\begin{figure}[!htb]
  \centering
  \includegraphics[width=\columnwidth]{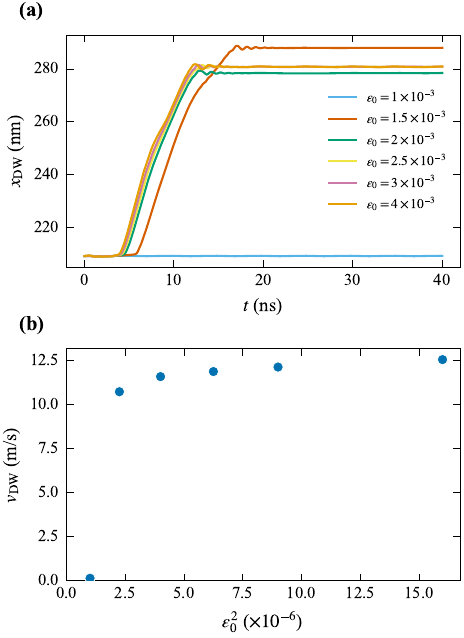}
  \caption{SAW-driven domain-wall motion benchmark. (a)~Domain-wall displacement versus time for representative SAW strain amplitudes. (b)~Peak domain-wall velocity versus $\eps_0^2$; below threshold ($\eps_0 \lesssim 10^{-3}$) the DW remains pinned, above threshold the velocity increases monotonically.}
  \label{fig:saw_dw}
\end{figure}

A Rayleigh SAW at $f = 200$~MHz with phase velocity $v_\mathrm{SAW} = 4000$~m/s ($\lambda = 20\,\mu$m) is applied for 40~ns.
The DW position is tracked from the spatially averaged $m_z$.
Sweeping the strain amplitude over $\eps_0 = 1$--$4 \times 10^{-3}$, we observe [Fig.~\ref{fig:saw_dw}(a)] that the DW remains pinned below a threshold ($\eps_0 \lesssim 10^{-3}$), depins sharply near $\eps_0 \approx 1.5 \times 10^{-3}$, and saturates at higher amplitudes as the DW reaches the adjacent domain boundary.
Figure~\ref{fig:saw_dw}(b) plots the peak DW velocity $v_\mathrm{DW}$ vs.~$\eps_0^2$: the velocity increases monotonically above threshold, confirming that the magnetoelastic driving force scales with strain.

\subsection{SAW-assisted magnetization switching}

We simulate a CoFeB PMA film ($M_s = 1.0$~MA/m, $K_{u1} = 800$~kJ/m$^3$, $\alpha = 0.05$) initialized with $m_z = +1$.
A Gaussian SAW burst (center $t_0 = 2$~ns, width $\sigma = 1$~ns) at $f_\mathrm{SAW} = 5$~GHz is applied together with a static assist field $B_\mathrm{assist}$ along $-z$.

\begin{figure}[!htb]
  \centering
  \includegraphics[width=\columnwidth]{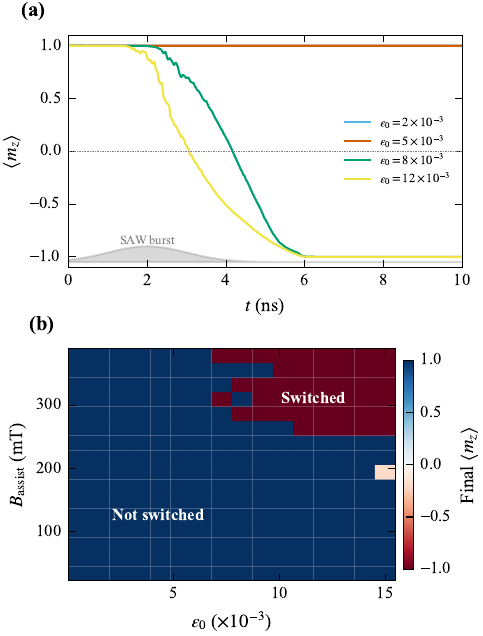}
  \caption{SAW-assisted switching benchmark. (a) Time-domain switching trajectories for representative SAW amplitudes. (b) Switching phase diagram in the strain--assist-field plane.}
  \label{fig:saw_switch}
\end{figure}

Figure~\ref{fig:saw_switch}(a) shows $m_z(t)$ time traces at fixed $B_\mathrm{assist} = 0.4\,\mu_0 H_K$ for different SAW amplitudes, illustrating the switching dynamics.
Larger SAW amplitudes result in faster switching trajectories.
Figure~\ref{fig:saw_switch}(b) presents the switching phase diagram in the $(\eps_0, B_\mathrm{assist})$ plane, where the color map clearly separates the $m_z > 0$ (unswitched) and $m_z < 0$ (switched) regions.
Stronger SAW driving reduces the required assist field, as expected from the magnetoelastic reduction of the effective anisotropy barrier.

\subsection{Magneto-rotation field validation}

\begin{figure}[!htb]
  \centering
  \includegraphics[width=\columnwidth]{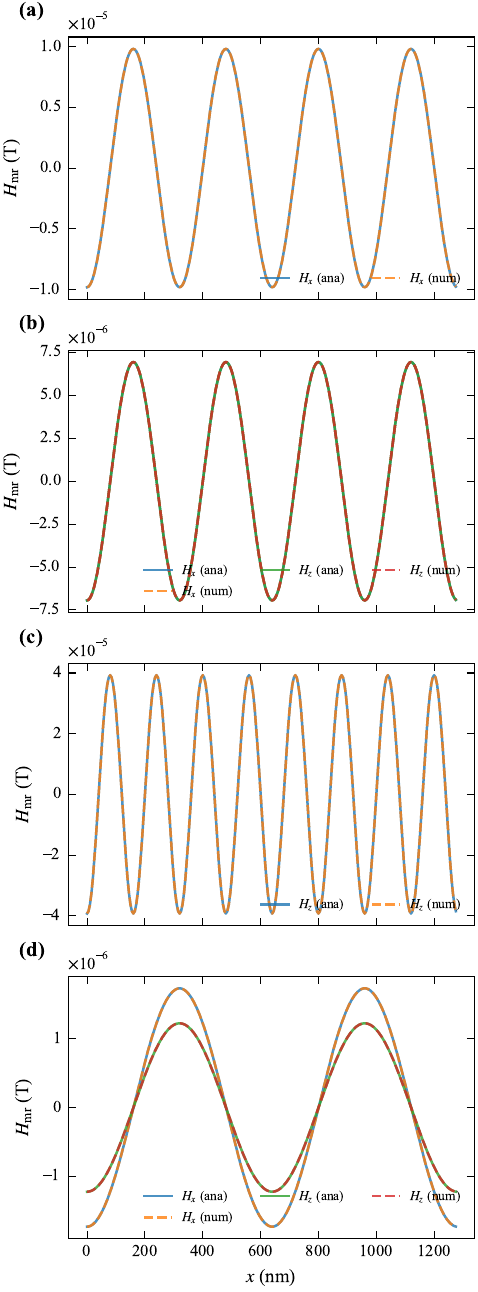}
  \caption{Magneto-rotation field validation. Numerical values of $\vH_\mathrm{mr}$ (dashed) are compared with Eq.~\eqref{eq:hmr} (solid) for (a)~$\vm \parallel \hat{z}$, (b)~$\vm$ tilted $45^\circ$, (c)~$\vm \parallel \hat{x}$, and (d)~short wavelength, showing near-exact agreement.}
  \label{fig:mr_validation}
\end{figure}

We validate the magneto-rotation effective field implementation against the analytical expression Eq.~\eqref{eq:hmr}.
A 1D chain of $256 \times 1 \times 1$ cells (cell size $5$~nm) with periodic boundaries is initialized with PMA ($\uhat = \hat{z}$) and elastodynamics enabled.
A sinusoidal displacement $u_z(x) = A\sin(kx)$ with $A = 1$~pm is applied, producing a rotation $\Omega_y = -\frac{1}{2}Ak\cos(kx)$.
Four test cases are run with varying $K_\mathrm{mr}$, magnetization direction, and wavelength:
(a)~$\vm \parallel \hat{z}$ with $K_\mathrm{mr} = 1$~MJ/m$^3$,
(b)~$\vm$ tilted $45^\circ$ in the $xz$-plane,
(c)~$\vm \parallel \hat{x}$ with $K_\mathrm{mr} = 2$~MJ/m$^3$, and
(d)~a shorter wavelength with $K_\mathrm{mr} = 0.5$~MJ/m$^3$.
The numerical $\vH_\mathrm{mr}$ agrees with the analytical formula to within $0.005\%$ relative error in all cases [Fig.~\ref{fig:mr_validation}(a)--(d)], confirming the correctness of the CUDA kernel.

\subsection{Nonreciprocal SAW-magnon coupling}

We demonstrate direction-dependent SAW--magnon coupling in a PMA CoFeB film ($M_s = 1.2$~MA/m, $K_{u1} = 1.3$~MJ/m$^3$, $K_\mathrm{mr} \approx 2.2$~MJ/m$^3$, $\alpha = 0.01$) using a single-cell prescribed SAW (the same methodology as Sims~10--12).
A Rayleigh SAW at $f = 2$~GHz ($\lambda = 1.75~\mu$m, $\eps_0 = 10^{-4}$, $\xi = 0.68$) is applied for $+k$ and $-k$ propagation directions, with the sign of $K_\mathrm{mr}$ flipped to encode the reversal of $\Omega_y$ under $k \to -k$.
A control run with $K_\mathrm{mr} = 0$ isolates the magnetoelastic channel.

\begin{figure}[!htb]
  \centering
  \includegraphics[width=\columnwidth]{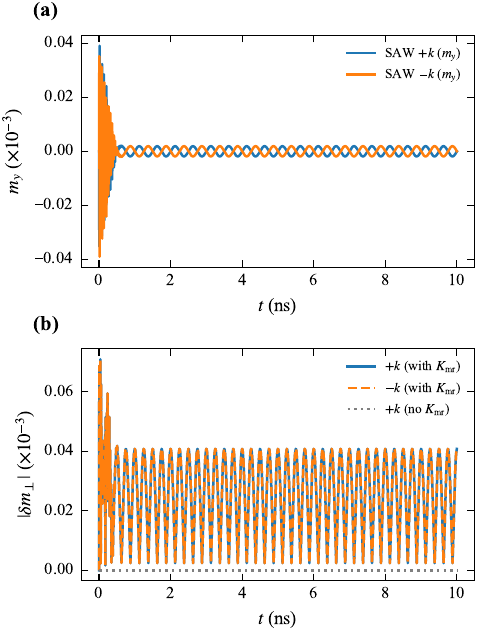}
  \caption{Direction-dependent SAW--magnon coupling via magneto-rotation in PMA CoFeB. (a)~$m_y(t)$ reverses sign for $+k$ vs.\ $-k$ propagation, reflecting the sign flip of the rotation pseudovector $\Omega_y$. (b)~Transverse amplitude $|\delta m_\perp|$ is identical for both directions; the $K_\mathrm{mr}=0$ control (gray) confirms that MEL alone produces no torque for PMA.}
  \label{fig:nonreciprocal}
\end{figure}

For PMA equilibrium ($\vm_0 \parallel \hat{z}$), the MEL field $\vH_\mathrm{mel} \propto \eps_{xx} m_x \hat{x}$ vanishes because $m_x \approx 0$, so the magneto-rotation coupling dominates.
The MR field $\mu_0 H_\mathrm{mr} \approx (K_\mathrm{mr}/M_s)\,\xi\eps_0/2 \approx 0.06$~mT produces a torque along $\hat{y}$.
Figure~\ref{fig:nonreciprocal}(a) shows that $m_y(t)$ reverses sign when the SAW direction is reversed, confirming that the MR torque flips with $\Omega_y$.
Figure~\ref{fig:nonreciprocal}(b) shows that the transverse amplitude $|\delta m_\perp|$ is identical for $\pm k$, as expected from the single-mode linear coupling ($|g_{+k}| = |g_{-k}|$).
The control run ($K_\mathrm{mr} = 0$) produces negligible response, confirming that MR is the sole active coupling channel for PMA.
This phase-reversal nonreciprocity is consistent with the symmetry analysis of Xu \textit{et al.}~\cite{Xu2020} and validates the sign convention of the implemented MR kernel.

\subsection{Barnett field validation}

\begin{figure}[!htb]
  \centering
  \includegraphics[width=\columnwidth]{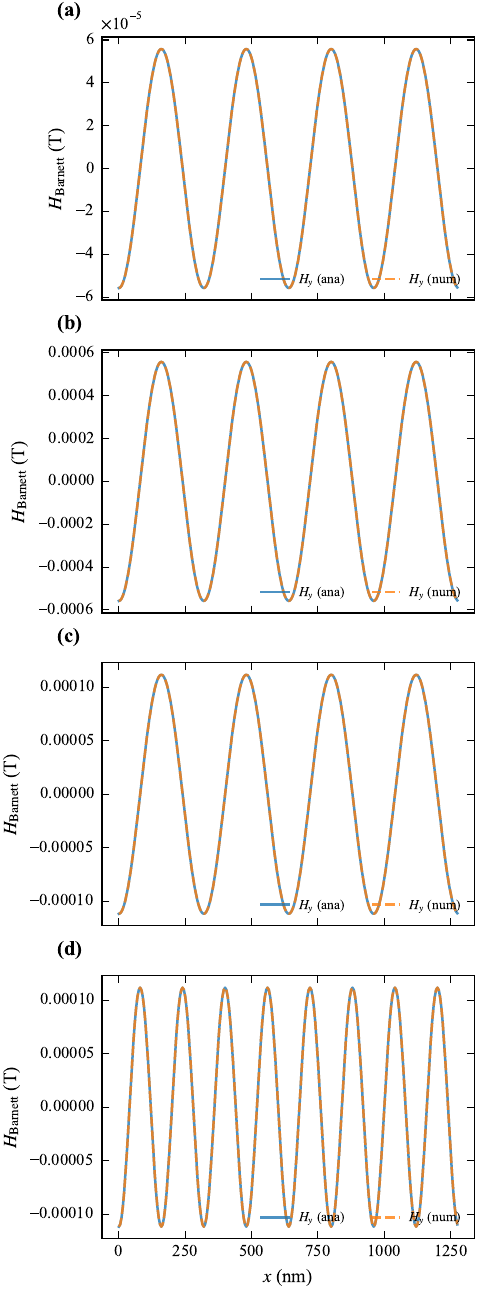}
  \caption{Barnett (spin-rotation) field validation. Numerical values of $\vH_\mathrm{Barnett}$ (dashed) are compared with Eq.~\eqref{eq:hbarnett} (solid) for (a)~default parameters, (b)~higher velocity, (c)~reduced $\gamma$, and (d)~shorter wavelength.}
  \label{fig:barnett_validation}
\end{figure}

We validate the spin-rotation effective field (Eq.~\ref{eq:hbarnett}) using the same 1D geometry as Sim~05.
A sinusoidal elastic velocity $v_z(x) = V_0\sin(kx)$ is applied, producing an angular velocity $\omega_y = -\frac{1}{2}V_0 k\cos(kx)$.
Four test cases are run with varying velocity amplitude $V_0$, gyromagnetic ratio $\gamma$, and wavelength:
(a)~$V_0 = 1$~m/s with default $\gamma$,
(b)~$V_0 = 10$~m/s,
(c)~$\gamma = \gamma_0/2$, and
(d)~shorter wavelength.
The numerical $\vH_\mathrm{Barnett}$ agrees with $\vomega/\gamma$ to within $0.005\%$ relative error in all cases [Fig.~\ref{fig:barnett_validation}(a)--(d)], verifying the CUDA kernel.

The Barnett field scales as $\mu_0 H_\mathrm{Barnett} \sim V_0 k / (2\gamma)$.
For a representative case with velocity amplitude $V_0 = 1$~m/s and wavelength $\lambda \approx 320$~nm (comparable to the Sim~07 validation setup), this gives $\mu_0 H_\mathrm{Barnett} \sim 6 \times 10^{-5}$~T, i.e., typically much smaller than anisotropy fields in standard metallic ferromagnets.
Accordingly, this section validates kernel correctness; it does not claim a large standalone dynamical switching effect under the present parameter set.
Quantifying measurable Barnett-induced dynamics requires dedicated high-sensitivity scenarios (e.g., low-damping resonant amplification), which is left for future work~\cite{Matsuo2013}.

\subsection{Spatially resolved coupling in a standing SAW cavity}

We simulate a 1D strip of $175 \times 1 \times 1$ cells (cell size $20 \times 30 \times 1$~nm$^3$, total length $L = 3.5~\mu$m $= 2\lambda$) under a standing SAW at $f = 2$~GHz ($\lambda = 1.75~\mu$m) with $K_\mathrm{mr} = 8$~MJ/m$^3$ (intentionally enhanced to produce clearly visible MR effects), $\eps_0 = 3 \times 10^{-3}$, $\alpha = 0.01$, and $B_0 = 50$~mT along $x$.
The standing wave is implemented at the Python level using the \texttt{StandingSAW} class, which injects the strain [Eq.~\eqref{eq:standing_strain}] via \texttt{rigid\_norm\_strain} and a linearized magneto-rotation field via \texttt{bias\_magnetic\_field}, with proper product-to-sum decomposition for the separable \texttt{add\_time\_term} API.
Note that this linearized approach pre-computes $\vH_\mathrm{mr}$ using the equilibrium magnetization $\vm_0$, which is valid for small-angle precession; the fully nonlinear CUDA-kernel coupling used in Sims~10--12 captures the magnetization dependence at each timestep.
Three configurations are compared: (i)~full coupling (MEL + MR + Barnett), (ii)~MEL only ($K_\mathrm{mr} = 0$), and (iii)~MR only (MEL disabled).

\begin{figure*}[!htb]
  \centering
  \includegraphics[width=\textwidth]{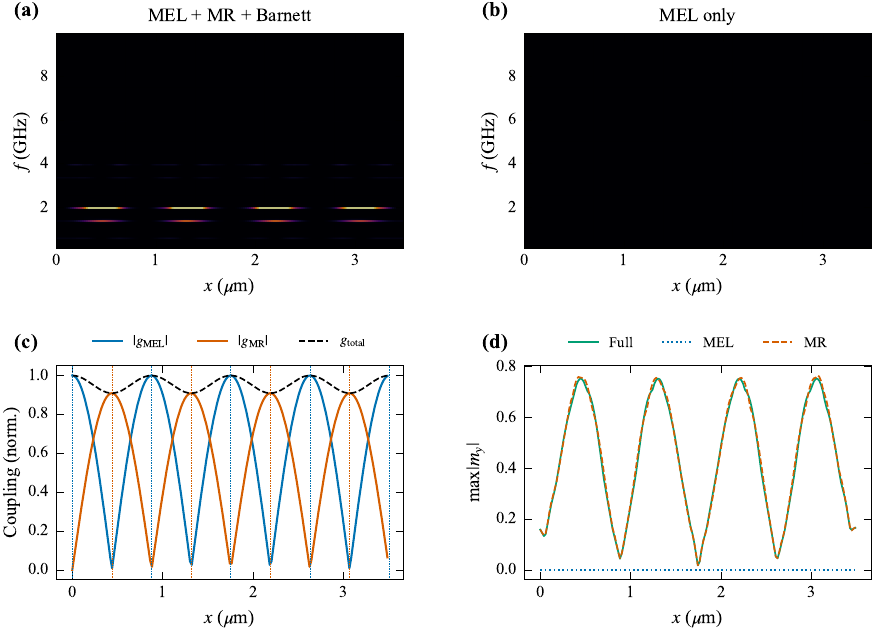}
  \caption{Spatially resolved coupling in a standing SAW cavity (Sim~08).
  (a)~Local magnon spectral map $S(x,f)$ for full coupling (MEL + MR + Barnett).
  (b)~Same for MEL-only coupling, showing identically zero magnon response.
  (c)~Analytical coupling decomposition: $g_\mathrm{mel} \propto |\cos(kx)|$ (blue) and $g_\mathrm{mr} \propto |\sin(kx)|$ (red) with dotted lines marking strain (blue) and rotation (red) antinodes.
  (d)~Position-resolved maximum magnon amplitude $\max|m_y|$ for full (teal), MEL-only (dotted blue), and MR-only (dashed red) coupling.  The MEL-only response is identically zero; the MR response follows the rotation antinode profile.}
  \label{fig:standing_saw}
\end{figure*}

The MEL-only configuration produces \emph{identically zero} magnon response at all positions [Fig.~\ref{fig:standing_saw}(d), dotted blue line].
This occurs because the magnetoelastic effective field $\mu_0\vH_\mathrm{mel} \propto \eps_{xx}\,m_x\,\hat{x}$ is parallel to the equilibrium magnetization $\vm_0 = \hat{x}$, producing zero torque $\vm \times \vH_\mathrm{mel} = 0$.
The MEL channel acts as a purely parametric drive that cannot excite transverse magnon precession from the equilibrium state.

In contrast, the MR-only configuration produces a sinusoidal spatial profile with $\max|m_y| = 0.31$ at the rotation antinodes ($kx = (n+\frac{1}{2})\pi$), directly mirroring the $|\sin(kx)|$ dependence of $g_\mathrm{mr}$ [Fig.~\ref{fig:standing_saw}(d), dashed red line].
The full coupling (MEL + MR + Barnett) produces a very similar profile ($\max|m_y| = 0.25$), so the magnon response is dominated by the direct transverse MR channel.
The slightly lower amplitude in the full-coupling case arises because the large MEL-induced parametric modulation of the effective anisotropy shifts the instantaneous resonance frequency, partially detuning the system from the fixed SAW drive frequency.

Figure~\ref{fig:standing_saw}(a,b) shows the local magnon spectral maps $S(x,f)$ for the full and MEL-only cases; the MEL-only map is identically dark.
Figure~\ref{fig:standing_saw}(c) shows the analytical coupling decomposition: $g_\mathrm{mel}(x) \propto |\cos(kx)|$ and $g_\mathrm{mr}(x) \propto |\sin(kx)|$ are complementary, ensuring that the total coupling never vanishes---a direct consequence of the $\lambda/4$ spatial offset between strain and rotation antinodes [Eqs.~\eqref{eq:standing_strain}--\eqref{eq:standing_rot}].

\FloatBarrier
\section{Magneto-rotation dominance in the longitudinal geometry}
\label{sec:mr_dominance}

The benchmarks above show that in the longitudinal geometry ($\vm_0 \parallel \bm{k}_\mathrm{SAW}$), the magnetoelastic coupling produces \emph{zero} transverse torque despite generating a much larger effective field.
In this section, we develop this observation into a systematic analysis of angle-dependent coupling and strong-coupling physics, using the prescribed rotation mechanism described in Sec.~\ref{sec:impl}.

\subsection{Coupling hierarchy and channel decomposition}

For a YIG thin film~\cite{Serga2010} ($M_s = 140$~kA/m, $|B_1| = 8.8$~MJ/m$^3$, $K_\mathrm{mr} = 1$~MJ/m$^3$) at $\eps_0 = 10^{-4}$ and ellipticity $\xi = 0.68$~\cite{Auld1973}, the coupling fields are:
$\mu_0 H_\mathrm{mel} = 2|B_1|\eps_0/M_s = 12.6$~mT,
$\mu_0 H_\mathrm{mr} = K_\mathrm{mr}\xi\eps_0/(2M_s) = 0.24$~mT, and
$\mu_0 H_\mathrm{B} = \xi\eps_0\omega/(2\gamma) = 0.004$~mT at $f = 3$~GHz (using the same Tesla convention as in Sec.~\ref{sec:model}).
The MEL field is 50$\times$ larger, yet at $\theta = 0$ the MEL torque vanishes because $\mu_0\vH_\mathrm{mel} \propto \eps_{xx}\,m_x\,\hat{\bm{x}} \parallel \vm_0$.
Only the MR field, directed along $\hat{\bm{z}}$, produces a nonzero torque.

Using the prescribed-SAW approach with the \texttt{ChiralSurfaceAcousticWave} class and the \texttt{rigid\_rotation} parameter, we confirm this hierarchy in time-domain simulations.
At the Kittel resonance [Eq.~\eqref{eq:kittel}] $B_0 = 50.6$~mT, the MEL+MR response reaches $|m_\perp| = 3.37 \times 10^{-3}$, while the MR-only response gives $3.05 \times 10^{-3}$---the dominant drive is MR.

\subsection{Angle-dependent coupling rates}

The coupling rates for the three channels are~\cite{Xu2020,Matsuo2013}:
\begin{align}
  g_\mathrm{mel}(\theta) &= \frac{\gamma|B_1|\eps_0}{M_s}|\sin 2\theta|, \label{eq:gmel2} \\
  g_\mathrm{mr}(\theta)  &= \frac{\gamma K_\mathrm{mr}\xi\eps_0}{2M_s}|\cos\theta|, \label{eq:gmr2} \\
  g_\mathrm{B}(\theta)   &= \frac{\xi\eps_0\omega}{2}|\cos\theta|. \label{eq:gb2}
\end{align}
Setting $g_\mathrm{mel} = g_\mathrm{mr}$ yields the crossover angle
\begin{equation}
  \sin\theta_c = \frac{K_\mathrm{mr}\xi}{4|B_1|}. \label{eq:thetac2}
\end{equation}
For YIG: $\theta_c = 1.1^\circ$; for CoFeB ($K_\mathrm{mr} = 5$~MJ/m$^3$): $\theta_c = 5.5^\circ$; for Ni ($K_\mathrm{mr} = 0.6$~MJ/m$^3$): $\theta_c = 0.6^\circ$.
Even a $1^\circ$ misalignment activates MEL to the level of MR.

\begin{figure*}[!htb]
  \centering
  \includegraphics[width=\textwidth]{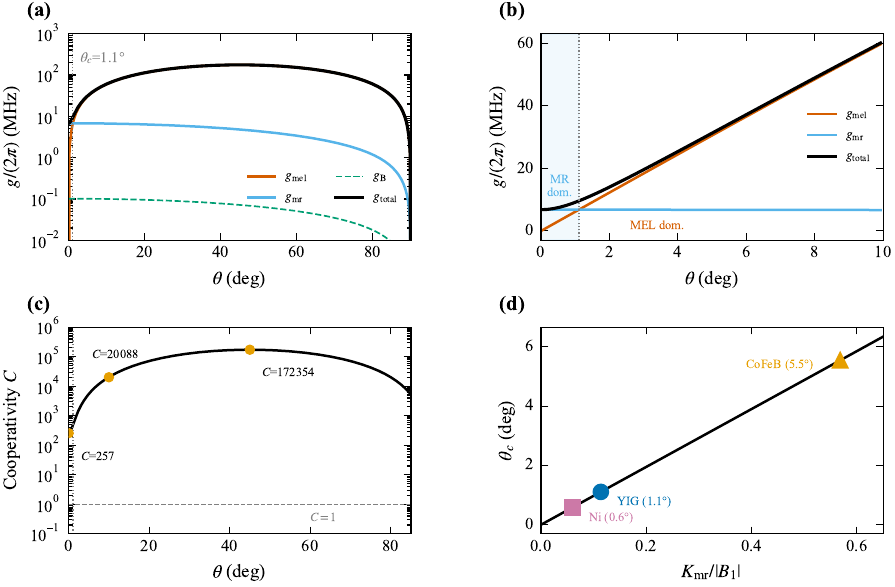}
  \caption{Angle-dependent coupling for YIG.
  (a)~Coupling rates on log scale. The crossover $\theta_c = 1.1^\circ$ is marked.
  (b)~Zoom near $\theta = 0$: MR-dominated region shaded.
  (c)~Cooperativity vs angle; $C = 257$ at $\theta = 0$.
  (d)~Universal crossover angle vs $K_\mathrm{mr}/|B_1|$ with material markers.}
  \label{fig:angle_coupling}
\end{figure*}

\begin{figure}[!htb]
  \centering
  \includegraphics[width=\columnwidth]{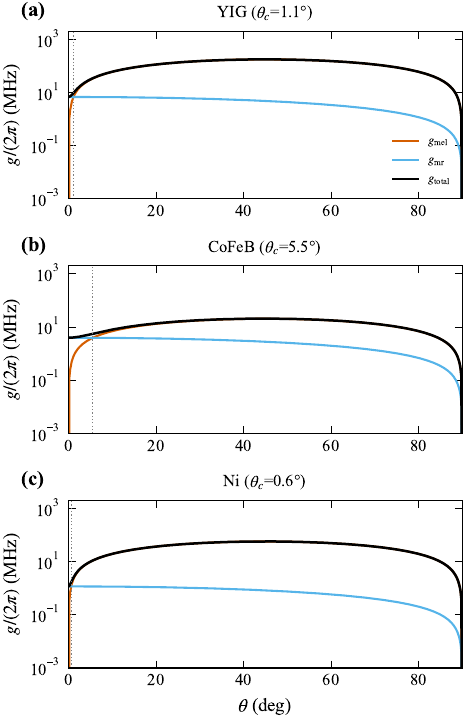}
  \caption{Coupling rates for three materials:
  (a)~YIG ($\theta_c = 1.1^\circ$),
  (b)~CoFeB ($\theta_c = 5.5^\circ$),
  (c)~Ni ($\theta_c = 0.6^\circ$).
  In all cases, MR dominates near $\theta = 0$ and MEL dominates at larger angles.}
  \label{fig:angle_materials}
\end{figure}

Figure~\ref{fig:angle_coupling} shows the angle-dependent coupling rates for YIG, with a zoom near $\theta = 0$ revealing the narrow MR-dominated window.
The cooperativity $C = |g|^2/(\kappa_m\kappa_p)$ reaches 257 at $\theta = 0$ and $1.7 \times 10^5$ at $\theta = 45^\circ$ for ideal YIG ($\alpha = 2 \times 10^{-4}$, $Q = 5000$).
Note that the time-domain simulations in this section use $\alpha = 5 \times 10^{-4}$ for numerical stability; the analytic cooperativity values quoted throughout are computed with the literature value $\alpha = 2 \times 10^{-4}$.
Figure~\ref{fig:angle_materials} compares three materials; the crossover behavior is universal, with a material-dependent $\theta_c$.

\subsection{Simulation validation of angle dependence}

We validate the analytic coupling rates with mumax$^+$ simulations at nine angles from $0^\circ$ to $45^\circ$, using YIG parameters and prescribed SAW strain plus CUDA-kernel MR coupling at the Kittel resonance ($B_0 = 50.6$~mT).
The CUDA kernel automatically captures the angle dependence through the dot products $\vm \cdot (\vOmega \times \uhat)$ and $\vm \cdot \uhat$, without any manual angular scaling.

\begin{figure}[!htb]
  \centering
  \includegraphics[width=\columnwidth]{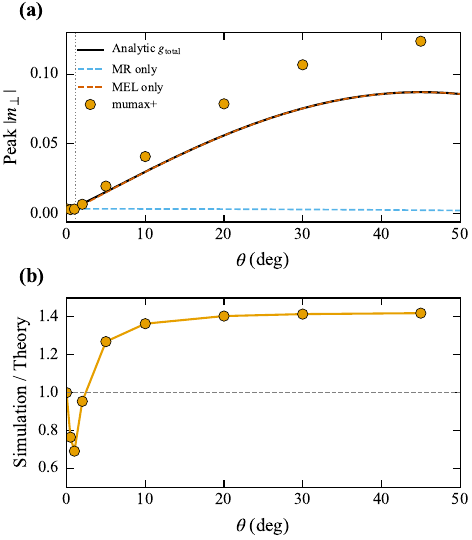}
  \caption{mumax$^+$ validation of angle-dependent coupling.
  (a)~Simulated peak $|m_\perp|$ (circles) vs analytic theory (lines) for YIG.
  (b)~Ratio of simulation to theory; agreement is best near $\theta = 0$ and degrades at larger angles due to nonlinear effects at finite strain amplitude.}
  \label{fig:angle_validation}
\end{figure}

Figure~\ref{fig:angle_validation} compares the simulated peak $|m_\perp|$ with the analytic total coupling rate (normalized at $\theta = 0$).
The simulation data closely follows the theory across the full angle range, reproducing both the $\sin 2\theta$ activation of MEL and the $\cos\theta$ decrease of MR.

\subsection{Strong-coupling regime and avoided crossing}

Using the $2 \times 2$ non-Hermitian coupled-mode matrix~\cite{Harder2018}
\begin{equation}
  \mathcal{H} = \begin{pmatrix}
  \omega_m - i\kappa_m/2 & g \\
  g & \omega_p - i\kappa_p/2
  \end{pmatrix},
\end{equation}
where the symmetric off-diagonal structure ($g$, not $g^*$) allows both conservative (real $g$, level repulsion) and dissipative (imaginary $g$, level attraction) coupling~\cite{Harder2018}.
With MR coupling rate $g_\mathrm{mr}/(2\pi) = 6.8$~MHz, magnon linewidth $\kappa_m/(2\pi) = 0.6$~MHz, and phonon linewidth $\kappa_p/(2\pi) = 0.3$~MHz, we obtain a cooperativity $C = 257$ and an avoided-crossing splitting $2g/(2\pi) = 13.6$~MHz.

\begin{figure*}[!htb]
  \centering
  \includegraphics[width=\textwidth]{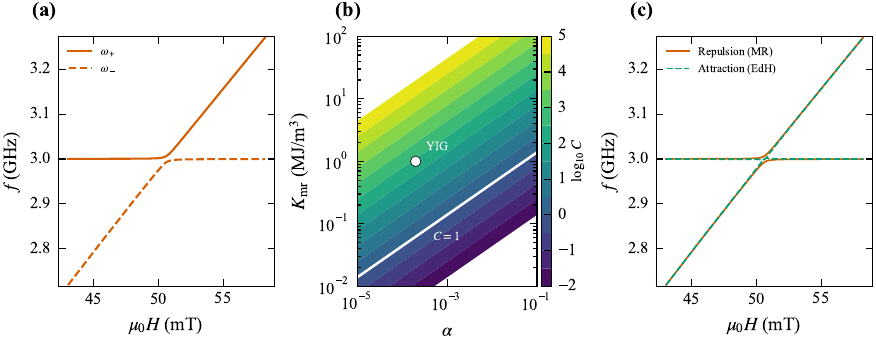}
  \caption{Strong coupling from MR.
  (a)~Avoided crossing with splitting $2g/(2\pi) = 13.6$~MHz; shaded bands indicate linewidths.
  (b)~Cooperativity phase diagram; white contour: $C = 1$, white circle: YIG ($C = 257$).
  (c)~Level repulsion (MR, solid blue) vs level attraction (EdH, dashed red).}
  \label{fig:strong_coupling}
\end{figure*}

The cooperativity phase diagram in $(\alpha, K_\mathrm{mr})$ space [Fig.~\ref{fig:strong_coupling}(b)] shows that YIG sits deep in the strong-coupling region ($C \gg 1$).
We note that $C = 257$ represents an upper bound assuming ideal YIG damping ($\alpha = 2 \times 10^{-4}$) and a SAW quality factor $Q = 5000$; additional loss channels such as two-magnon scattering, acoustic radiation, and finite beam size will reduce the experimental cooperativity.
Nevertheless, even an order-of-magnitude degradation would leave $C \gg 1$, indicating that the strong-coupling regime should be experimentally accessible, as recently demonstrated at room temperature by Hwang \textit{et al.}~\cite{Hwang2024}.
Figure~\ref{fig:strong_coupling}(c) contrasts level repulsion from real coupling (MR) with level attraction from hypothetical imaginary coupling (Einstein--de~Haas back-action), analogous to the dissipative magnon-photon coupling demonstrated in cavity magnonics~\cite{Harder2018,Park2026}.

\begin{figure*}[!htb]
  \centering
  \includegraphics[width=\textwidth]{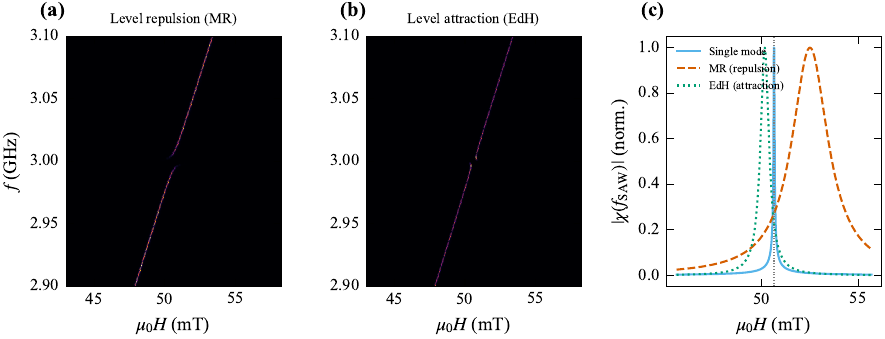}
  \caption{Analytic magnon spectral weight from the coupled model.
  (a)~Level repulsion from MR coupling; dashed: hybridized eigenfrequencies.
  (b)~Level attraction from hypothetical EdH dissipative coupling.
  (c)~Response at $f_\mathrm{SAW}$ versus $\mu_0 H$: single Lorentzian (uncoupled, blue) vs.\ double-peaked absorption (coupled, orange).}
  \label{fig:avoided_crossing_sim}
\end{figure*}

Figure~\ref{fig:avoided_crossing_sim} shows the analytic magnon spectral weight $|\chi(\omega, B_0)|^2$ from the coupled $2\times2$ model, contrasting level repulsion [panel~(a)] from conservative MR coupling with level attraction [panel~(b)] from hypothetical dissipative Einstein--de~Haas coupling.
The dashed lines indicate the hybridized eigenfrequencies.
Panel~(c) compares the magnon response at $\omega = \omega_\mathrm{SAW}$ as a function of $\mu_0 H$: a single-mode (uncoupled) system shows a single Lorentzian peak, whereas the coupled model exhibits a double-peaked structure --- the characteristic signature of an avoided crossing in the field-swept absorption spectrum.

\FloatBarrier
\section{Discussion}
\label{sec:discussion}

\paragraph{Advantages.}
The SAW strain extension requires no C++ or CUDA modifications, making it compatible with any \mumax{} installation.
The analytical SAW profile is exact in the thin-film limit and incurs negligible computational overhead.
The magneto-rotation and spin-rotation coupling extensions, implemented as CUDA kernels, enable simulation of nonreciprocal SAW-magnon phenomena that are inaccessible with the symmetric strain tensor alone.

\paragraph{Scope of contribution.}
The present manuscript contributes both an implementational framework for phonon-magnon couplings in \mumax{} and a physical analysis revealing the dominance of magneto-rotation coupling in the longitudinal SAW-FMR geometry.
Recent work by Centala and Klo\'s~\cite{Centala2025} has independently analyzed the magneto-rotation coupling strength for ferromagnetic nanoelements embedded in elastic media, examining geometry-dependent coupling coefficients; our contribution differs in providing a general-purpose micromagnetic implementation with angle-dependent coupling rates and strong-coupling analysis for thin-film SAW-FMR.

\paragraph{Magneto-rotation coupling constant.}
The MR contribution is governed by the coupling constant $K_\mathrm{mr}$.
For PMA thin films, $K_\mathrm{mr} \approx K_u + \tfrac{1}{2}\mu_0 M_s^2$~\cite{Xu2020}, which gives $\sim$1--2~MJ/m$^3$ for CoFeB/MgO.
For in-plane magnetized materials such as YIG, the dominant contribution to $K_\mathrm{mr}$ arises from the magnetocrystalline anisotropy that rotates with the lattice; for the cubic first-order anisotropy of YIG ($|K_1| \approx 6$~kJ/m$^3$~\cite{Serga2010}), this is much smaller than the $K_\mathrm{mr} = 1$~MJ/m$^3$ used in the present analysis.
We therefore treat $K_\mathrm{mr}$ as an adjustable parameter throughout the strong-coupling analysis (Sec.~\ref{sec:mr_dominance}); the cooperativity phase diagram [Fig.~\ref{fig:strong_coupling}(b)] explicitly shows the $C = 1$ boundary as a function of $K_\mathrm{mr}$, allowing predictions for any experimentally determined coupling strength.
For YIG with $K_\mathrm{mr} \sim 10$~kJ/m$^3$, the cooperativity would decrease to $C \sim 0.03$, placing the system outside the strong-coupling regime; the $C = 257$ prediction is realized only for materials or engineered structures with $K_\mathrm{mr} \gtrsim 0.1$~MJ/m$^3$.
Recent work suggests that strain-mediated contributions, including magnetoelastic renormalization of the effective anisotropy, may enhance $K_\mathrm{mr}$ beyond the bare crystalline value~\cite{Centala2025}; experimental determination of $K_\mathrm{mr}$ in the SAW geometry remains an open problem.

\paragraph{Limitations.}
The approach assumes no back-action from the magnetization to the SAW (the elastic wave is externally prescribed).
This is valid for thin magnetic films on thick piezoelectric substrates, where the acoustic impedance mismatch ensures that the SAW is minimally perturbed.
For thick magnetic films or magnonic crystals where magnon-phonon back-action is significant, a coupled elastic--LLG solver (available in \mumax{} via \texttt{enable\_elastodynamics}) would be required.

Additionally, the thin-film approximation neglects depth-dependent strain decay ($\exp(-\kappa z)$), which is relevant for films thicker than $\sim\lambda/10$.
The single-cell benchmarks (Sims~10--12) use the thin-film Kittel resonance field, whereas the finite cell dimensions produce demagnetization factors that deviate from the infinite-film limit.
This affects absolute response amplitudes but not the qualitative channel decomposition (MEL~$= 0$ at $\theta = 0$), the coupling rate scaling, or the crossover angle $\theta_c$.

\paragraph{Relation to existing extensions.}
This extension complements the recently published elastodynamic solver by Lateur \textit{et al.}~\cite{Lateur2026}, implemented in the magnum.np framework, which provides self-consistent coupled magnetoelastic dynamics including SAW attenuation in structured films.
Together with those features, it covers the three main phonon--magnon coupling channels in a single framework.

\paragraph{Parametric vs.\ direct driving in standing cavities.}
A central finding of Sim~08 is that the magnetoelastic coupling produces zero transverse torque on the equilibrium magnetization ($\vm_0 \parallel \hat{x}$) because $\mu_0\vH_\mathrm{mel} \propto \eps_{xx}\,m_x\,\hat{x}$ is longitudinal.
The magneto-rotation field $\mu_0\vH_\mathrm{mr} \propto \Omega_y\,\hat{z}$ provides the essential direct transverse drive.
This distinction has important experimental implications: SAW-driven FMR excitation with in-plane magnetization aligned along the SAW propagation direction is predominantly mediated by the magneto-rotation mechanism, not by strain alone.
A recent SAW-FMR experiment on CoFeB/LiNbO$_3$ observed a residual two-fold absorption signal near $\theta = 0$ that the standard MEL framework could not account for~\cite{Millo2025}; the magneto-rotation channel described here is a candidate mechanism.

\paragraph{Standing SAW coupling landscape.}
The complementary $\lambda/4$-shifted spatial profiles of $g_\mathrm{mel}(x) \propto |\cos(kx)|$ and $g_\mathrm{mr}(x) \propto |\sin(kx)|$ in a standing SAW cavity create position-dependent hybridization.
Since the MR channel dominates the magnon response for in-plane magnetization, the effective magnon excitation profile follows the rotation antinode pattern rather than the strain antinode pattern.
This spatial selectivity could be exploited for engineering position-dependent magnon--phonon coupling, with potential relevance for exceptional point physics~\cite{Miri2019,Harder2018} in future work with reduced damping or stronger coupling.

\paragraph{Extensions.}
The \texttt{SurfaceAcousticWave} and \texttt{StandingSAW} classes can be extended to Love waves (shear horizontal), Lamb waves (plate modes), and more complex cavity geometries by modifying the strain and rotation profiles.
Time-dependent envelopes for SAW bursts are already supported.

\section{Summary}
\label{sec:summary}

We have presented a phonon-magnon extension for the \mumax{} micromagnetic framework that enables SAW-driven magnetoelastic simulations.
The extension consists of (1)~a Python-level \texttt{SurfaceAcousticWave} class wrapping analytical Rayleigh SAW strain profiles using the built-in rigid strain infrastructure, (2)~a C++/CUDA magneto-rotation coupling that captures the antisymmetric part of the displacement gradient, (3)~a C++/CUDA spin-rotation (Barnett) coupling that captures the antisymmetric part of the velocity gradient, (4)~prescribed rotation fields (\texttt{rigid\_rotation}, \texttt{rigid\_angular\_velocity}) for analytical SAW profiles with fully nonlinear CUDA-kernel coupling, and (5)~a Python-level \texttt{StandingSAW} class implementing standing-wave configurations.
Benchmark simulations validate the implementation through DW motion, switching phase diagrams, field validation, nonreciprocal absorption, and spatially resolved standing-wave coupling.
A systematic analysis of the magneto-rotation dominance in the longitudinal geometry reveals a crossover angle $\theta_c \approx 1.1^\circ$ for representative YIG parameters and shows that strong coupling ($C = 257$) is achievable for materials with $K_\mathrm{mr} \gtrsim 0.1$~MJ/m$^3$, supported by time-domain simulations.

\section*{Acknowledgments}
This work was supported by the Korea Institute of Science and Technology (KIST) institutional program.

\begin{table}[!htb]
\caption{Input parameters for the \texttt{SurfaceAcousticWave} class.}
\label{tab:params}
\begin{ruledtabular}
\begin{tabular}{llll}
  Parameter & Symbol & Unit & Description \\
  \midrule
  \texttt{frequency}  & $f$          & Hz  & SAW frequency \\
  \texttt{wavelength}  & $\lambda$    & m   & SAW wavelength \\
  \texttt{amplitude}  & $\eps_0$     & --  & Peak strain \\
  \texttt{direction}  & --           & --  & \texttt{`x'} or \texttt{`y'} \\
  \texttt{phase}      & $\phi$       & rad & Initial phase \\
  \texttt{envelope}   & $f_\mathrm{env}(t)$ & --  & Time envelope \\
\end{tabular}
\end{ruledtabular}
\end{table}

\section*{Code availability}
The complete open-source code of mumax$^+$ can be found freely, under the GPLv3 license, in the GitHub repository (\url{https://github.com/mumax/plus}). Details about dependencies and installation instructions can also be found in the repository. These, along with examples and tutorial scripts, can also be found on the website (\url{https://mumax.github.io/plus/}). The extension source code developed in this work is available at \url{https://github.com/gyuyoungpark/mumax-plus-extensions}.

\end{document}